%% file: main.tex
\documentclass[12pt, hidelinks]{article}
\usepackage[small, bf]{caption}
\usepackage[utf8]{inputenc}
\usepackage{fullpage}
\usepackage{appendix}
\usepackage{graphicx}
\usepackage{amsmath}
\usepackage{amssymb}
\usepackage{amsthm, amsfonts}
\usepackage{url}
\usepackage{mathtools}
\usepackage{chngcntr}
\usepackage{nicefrac}
\usepackage{hyperref}
\usepackage{subdepth}
\usepackage{forest}
\usepackage{tikzducks}
\usetikzlibrary{shapes.geometric,positioning}
\usetikzlibrary{decorations.pathmorphing,shapes}
\tikzset{elli/.style={ellipse,draw}}
\newsavebox\Car
\newsavebox\Tree

\usepackage{comment}
\usepackage{bbm}

\usepackage[numbers]{natbib}  

\theoremstyle{definition}

\theoremstyle{theorem}

\counterwithin*{lemma}{subsection}

\usepackage{tikz}
\usepackage{subcaption}
\usetikzlibrary{arrows,automata,calc}

\tikzset{
  treenode/.style = {align=center, inner sep=0pt, text centered,
    font=\sffamily},
  arn_n/.style = {treenode, circle, white, font=\sffamily\bfseries, draw=black,
    fill=black, text width=1.5em},
  arn_r/.style = {treenode, circle, red, draw=red, 
    text width=1.5em, very thick},
  arn_x/.style = {treenode, rectangle, draw=black,
    minimum width=0.5em, minimum height=0.5em}
}

\input{defs.tex}

\title{Execution Welfare Across Solver-based DEXes}
\author{
   Yuki Yuminaga\\ Tenbin Labs \\\texttt{\small yuki@tenbinlabs.xyz}  \and
    Dex Chen\\ Semantic Layer \\\texttt{\small dex.c@semanticlayer.io} \and
    Danning Sui\\ Flashbots \\\texttt{\small danning@flashbots.net}
}

\begin{document}

\maketitle

\begin{abstract}
Decentralized exchanges (DEXes) have evolved dramatically since the introduction of Automated Market Makers (AMMs). In recent years, solver-based protocols have emerged as an alternative venue aiming to introduce competition for routing, access to offchain liquidity, and thereby improve end-user execution. Currently, these solver auctions are hosted on opaque backends, and the extent of price improvement they provide to end users remains unclear.

We conduct an empirical study of the \textit{execution welfare} that these protocols bring to users by analyzing data across different asset profiles (USDC-WETH and PEPE-WETH). Our results indicate that, compared to vanilla routing through Uniswap V2 or V3, solver-based protocols effectively enhance execution welfare for end users on DEXes within certain trade size ranges. This effect is most pronounced with USDC-WETH, a short-tail asset, and somewhat less significant with PEPE-WETH, a long-tail asset.

Additionally, we identify execution welfare discrepancies across solver-based platforms (e.g., CoWSwap, 1inchFusion, UniswapX), revealing potential inefficiencies due to solver market structure, variations in liquidity profile and inventory depth among solvers. These insights highlight both the advantages and challenges of solver-based trading, underscoring its role in improving execution outcomes while raising concerns about market concentration and competition dynamics.

\end{abstract}

\section{Introduction}
Many decentralized exchange (DEX) protocols have begun to leverage solvers (also called fillers or resolvers), who are third parties that compete for better execution of end-user trades in recent years. The most widely adopted solver-based DEXes include CoWSwap~\cite{cowfaq}, 1inchFusion~\cite{1inch2022}, and UniswapX~\cite{adams2023}. Despite the efforts that have gone into building solver-based DEXes, there has yet to be a comprehensive study of their impact on the trade execution of both long-tail and short-tail tokens compared to trade execution through Automated Market Makers (AMMs), such as Uniswap V2~\cite{uniswap2020} and V3~\cite{uniswap2021}.

The term \textit{solver} typically refers to third parties who run an algorithm or a set of algorithms that dynamically determine the best trade route for end users, leveraging onchain liquidity or offchain inventory to execute trades. Solvers can be broadly categorized into two types: \textit{AMM-solvers}, who aim to route across onchain liquidity with minimal gas costs and the maximal amount of token output for end users, and \textit{PMM-solvers}, who are professional market makers (PMMs) responding with quotes from private inventory. PMM-solvers typically hold inventory both on DEXes and Centralized Exchanges (CEXes) and hedge using profit-taking trading strategies. These market makers have been quoting in DEX routing behind request-for-quote (RFQ) systems on DEX aggregators like 0x~\cite{0xprotocol} and 1inch~\cite{1inch} before solver-based platforms emerged. Hence, RFQ traditionally refers to market maker quotes for DEX aggregator trades, until solver-based DEXes appeared and generalized it into a broader term. For example, UniswapX describes\cite{uniswap2024} their solver auction as an RFQ system. Furthermore, AMM-solvers can leverage RFQ protocols such as Hashflow~\cite{hashflow} to access offchain liquidity from professional market makers as well, effectively becoming hybrid solvers. In essence, solver-based DEXes aim to equip end users with access to a more diverse set of liquidity from external market makers and achieve better trade routing.

We initiate an empirical study of solver-based DEXes and study their price competitiveness and the effectiveness of their associated auctions to answer the following main research question: \textbf{How do solver-based DEXes compare to traditional AMMs in terms of user welfare?}
To measure this, we introduce the notion of \emph{execution welfare}, a metric used to assess the additional percentage of output assets that users receive via a solver-based DEX compared to naive routing that sources only from single AMM pool liquidity. We evaluate the execution welfare of major solver-based DEXes for two representative asset pairs (USDC-WETH and PEPE-WETH pairs), using their respective Uniswap V2 and V3 pools with largest liquidity as the baseline comparison. We discuss numerous factors influencing trade execution, such as solver competition, token incentives, trade size, market volatility, and gas costs. Our main research contributions are summarized below.

\subsection{Contributions}
\begin{itemize}
    \item We build a rich dataset, including onchain execution results across 3 major platforms and Binance orderbook prices for 2 asset pairs over a period of 6 months.
    \item We introduced the metric called \emph{execution welfare} and propose a framework to evaluate trade execution quality across solver-based DEXes. Our approach establishes the robust methodology for decoding onchain execution outcomes and defines a counterfactual benchmark using simulation tools.
    \item We demonstrate that execution welfare varies depending on trade sizes, asset profiles and benchmarking against Uniswap V2 or V3. We discover that market volatility does not show a significant correlation with the improved execution welfare of solver-based DEXes.
    \item We demonstrate that the differing liquidity landscape between short- and long-tail assets leads to variations in execution welfare. Additionally, we observe that liquidity profiles within solver sets vary across venues, trade sizes, and asset pairs, which in turn impacts execution welfare.
\end{itemize}

\section{Literature Review}
AMMs were introduced to address liquidity pricing challenges for long-tail assets, as they eliminate the need for market makers to actively quote prices or hold inventories. Instead, a set of smart contracts fulfills this role, autonomously facilitating trades. The asset pair's price moves around a curve defined by a mathematical formula (usually \(xy=k\) or \(x+y=k\)) embedded in the pool contract, and the trade is filled with onchain inventory, which can be deposited by any liquidity provider (LP) permissionlessly in the smart contract. Onchain trade execution has been extensively studied, highlighting various challenges associated with AMMs - the simple pricing mechanism also introduces hidden costs, such as slippage, since the asset price constantly moves as each order takes the liquidity from the pool. Users' trade may receive a lower amount of output token in the trade execution, from what they saw in the quoted price from the trading interface~\cite{0x2022hidden}. In certain cases, the trade can even revert if the executed price exceeds the slippage threshold. Combined with Ethereum's gas costs, slippage can significantly degrade execution quality, particularly for smaller trades~\cite{adams2024mev}. Consequently, improving onchain trade execution has emerged as a critical focus for teams across the industry.

Researchers and developers have proposed alternative AMM designs to improve trading experiences. These models aim to enhance capital efficiency~\cite{abgaryan2023dfmm}, ensure sustainability for liquidity providers~\cite{adams2024amAMM, canidio2024batchAMM}, and increase execution speed~\cite{chen2024samm}. Although these designs often involve trade-offs, such as the use of price oracles and reliance on offchain entities, they serve as valuable attempts to improving trade execution.

Diversifying liquidity sources is another strategy to enhance trade execution. For instance, integrating offchain liquidity through request-for-quote (RFQ) systems allows professional market makers and solvers—collectively referred to as fillers on UniswapX—to provide liquidity held on both centralized exchanges (CEXes) and decentralized exchanges (DEXes). 
\begin{figure}[t]
    \centering
    \includegraphics[width=0.65\linewidth]{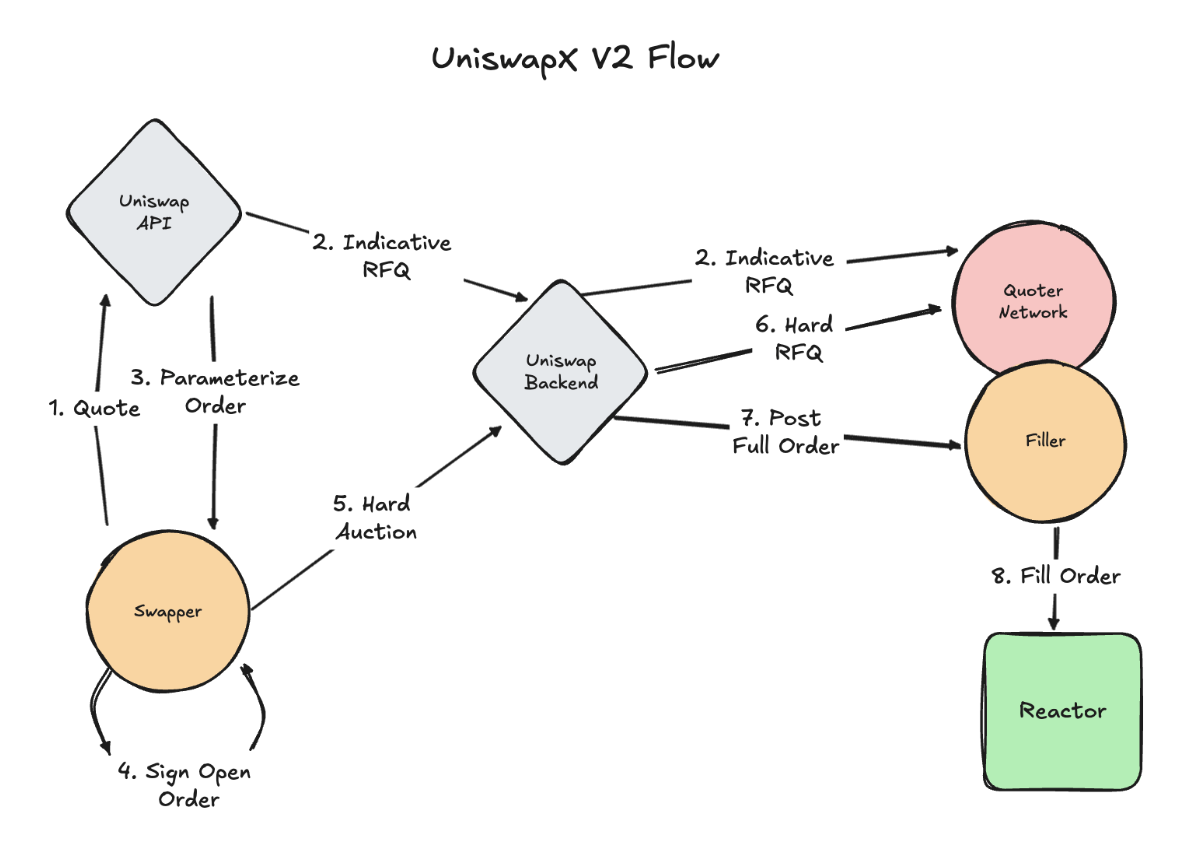}
    \caption{Trade flow for UniswapX V2.}
    \label{fig:unix-flow}
\end{figure}
Figure \ref{fig:unix-flow}, which depicts the UniswapX workflow, shows the RFQ process involves two stages: an initial \textit{indicative RFQ}, followed by a \textit{hard RFQ}. First, the Uniswap backend sends an indicative RFQ to the quoter network to gather preliminary price quotes without committing to a trade. Then, after the swapper finalizes order parameters and signs an open order, a hard RFQ is issued, this time requesting a firm quote from potential fillers. The filler with the best quote are allowed to execute the trade in the Reactor, ensuring competitive pricing, thereby improving execution welfare. An empirical study \cite{0x2023rfq} by 0x demonstrated that RFQ systems utilizing PMM liquidity enhance trade execution by offering more competitive pricing compared to AMMs. The study also introduced a liquidity model that explains why PMMs via RFQ favor and outperform at short-tail asset pairs and larger trade size, due to inventory risks and execution costs. More recent study ~\cite{chitra2024intent} shows that the inventory cost restrict new entry to the solver market and leads to oligopoly. Additionally, study shows that just-in-time (JIT) liquidity onchain can enhance trade execution~\cite{wan2022jit}, despite its potential negative impact on passive liquidity providers (LPs) and retail traders due to its adverse selection on uninformed flows~\cite{capponi2023jit}.
\begin{figure}
    \centering
    \includegraphics[width=1\linewidth]{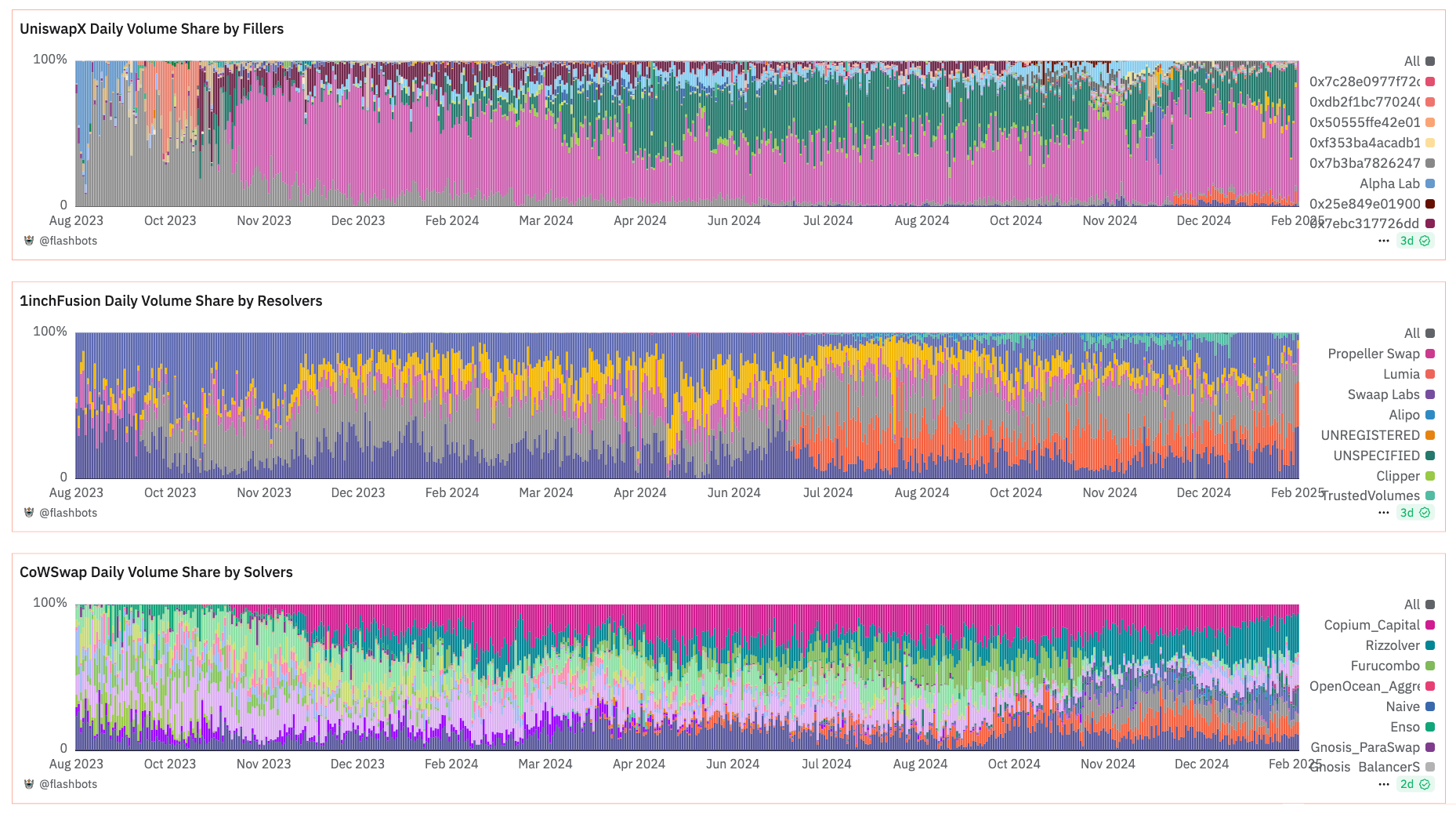}
    \caption{Solver Platforms Volume Share by Solvers, 2023 Aug to 2025 Feb}
    \label{fig:solver-volume}
\end{figure}

Figure \ref{fig:solver-volume} gives an overview of the current solver market structure across the three major platforms, with CoWSwap maintaining a more decentralized solver market structure. Among them, CoWSwap records an average daily trading volume of approximately \$200 million, while 1inchFusion totals around \$60 million and UniswapX around \$30 million.

On CoWSwap, the top one solver is a market maker (Wintermute) accounting for 25\% of the market share, and the next five leading solvers each hold approximately 10\%. 1inchFusion shows a greater level of centralization, where the top three solvers collectively control 80\% of total volume. This trend becomes even more pronounced in UniswapX, where just two market-making solvers, SCP and Wintermute, dominate over 90\% of total trading volume.


In examining price improvement and trade execution of DEXes with order flow auctions, a study was conducted on 1inchFusion and UniswapX, two of the three main solver-based DEXes. This research performed end-of-block simulations of counterfactual trades using the Uniswap’s universal router and mathematically decomposed the factors influencing execution welfare~\cite{bachu2024auction}. Another study~\cite{lladrovci2023cow} examined the impact of batch auctions on mitigating MEV and found that the occurrence of Coincidence of Wants (CoWs) on CoWSwap is negligible, offering limited benefits.

\section{Methodology}
Our study establishes a framework for a comprehensive analysis of execution welfare across all major solver-based DEXes, including CoWSwap. Using a top-of-block simulation result—where transaction outcomes are based on the final state of the previous block (i.e. the initial state of the current block)—we aim to eliminate any interactions between executed trades and counterfactual simulated trades to more accurately assess the execution welfare of solver-based DEXes against Uniswap V2 and V3. The core question we address is whether a solver-based approach can outperform V2 and V3 execution if they all operated under similar liquidity conditions within the same block.   

\subsection{Data Coverage for Executed Trades }
We collect and analyze data from block \texttt{18000000} to \texttt{19164987} on Ethereum mainnet, covering the six-month period from August 26, 2023, to February 5, 2024. Our dataset includes trades executed on UniswapX, 1inchFusion, and CoWSwap, with details on data collection provided in the Appendix.\ref{appendix:data-executed}.

\subsection{Execution Welfare}
To assess the execution quality of the solver-based protocols compared with AMM, we define the following metric:

\begin{equation}
\text{Execution Welfare} = \frac{\text{Solver output amount} - \text{AMM output amount after gas}}{\text{AMM output amount after gas}}
\end{equation}

The AMM output amount is calculated by simulating counterfactual trades based on the liquidity distribution of Uniswap V2 and V3 AMM pools and the top-of-block liquidity status for each solver-filled trade. Execution welfare shows the percentage of additional output assets the user received with solver-based order compared to the amount of output assets the user could have received with AMM in simulation, after accounting for gas fees and LP fees. The higher the execution welfare, the better the trade execution is. This metric can also be understood as a price improvement of solver-based DEXes similarly in~\cite{bachu2024auction}, with one caveat that the referenced study used an end-of-block simulation.

\subsubsection{Gas Estimation}
Tenderly~\cite{tenderly2024gas} was used to simulate gas units for swaps in the Uniswap V2 and V3 benchmarks, replicating the direction and amount of solver-filled orders.

To accurately apply the gas price for simulated swaps, the distribution of existing gas prices was analyzed. A comparison was made between an optimistic gas estimate and a more conservative estimate, with the latter being adopted as it more closely reflects the gas costs of solver-executed swaps. The total gas fees was then estimated by multiplying gas price and gas units. Further details are provided in Appendix~\ref{appendix:gas}.

\subsubsection{Uniswap V2 Simulation}
For each solver-based order filled in block number \(N\), the state of Uniswap V2 USDC-WETH pool (deployed at \texttt{0xB4e1...C9Dc}\cite{etherscanV2usdcpool}) at block number \(N-1\) is retrieved by calling the view functions on reserves from Infura API~\cite{infura}, with the block number specified. Since Infura API view function calls are executed at the end of the block, we call the view function at block \(N-1\) to assume top-of-block \(N\) execution of our simulated transactions, preventing any liquidity interference caused by the corresponding solver-based order.

If the simulated swap is WETH \(\rightarrow\) USDC, then the gas cost (in ETH) is converted to transaction fee (in USDC) by assuming the spot price as \(\frac{\text{reserve of USDC in Uniswap V2 pool}}{\text{reserve of ETH in Uniswap V2 pool}}\), and subtracted from the output amount (in USDC). Alternatively, if the simulated swap is USDC \(\rightarrow\) WETH, then the gas cost is directly deducted from the output amount, as both the transaction fee and the output amount are denoted in ETH.

The simulation also assumes that the AMM LP charges a 0.3\% fee on swaps, and if the output amount after gas fees is less than 0, the entire record (including the actual solver-based DEX order) is discarded and not included in the research. This happens occasionally with small-sized trades with high gas costs.

Same methodology was implemented for largest PEPE-WETH V2 pool (deployed at \texttt{0xA43f...Ec9f}\cite{etherscanV2pepepool}) with a fixed 0.3\% LP fee rate considered.

\subsubsection{Uniswap V3 Simulation}
In the simulation of Uniswap V3, we utilize the official TypeScript SDK provided by Uniswap V3. Interactions with the Uniswap V3 Quoter contract (deployed at \texttt{0xb273...5AB6}\cite{etherscanQuoter}) are conducted using the \texttt{quoteExactInputSingle} method. This method is called with the parameters \texttt{[tokenInAddress, tokenOutAddress, poolFee, amountIn, sqrtPriceLimitX96 (set to 0)]}, and the contract returns an estimated output amount, \texttt{amountOut} with pool's LP fees accounted. For each solver-based order filled at block number \(N\), the status of the largest Uniswap V3 USDC-WETH pool (deployed at \texttt{0x88e6...5640}\cite{etherscanV3usdcpool}) at block number \(N-1\) is retrieved by interacting with the Uniswap V3 pool contract, specifying the relevant block number. Same approach is applied for largest PEPE-WETH v3 pool (deployed at \texttt{0x1195...7b58} \cite{etherscanV3pepepool}).

This tool enables us to simulate the output amount from a single pool. Therefore, our primary objective is to establish a benchmark for comparing if solver-based design outperform vanilla AMM pool pricing, and welfare differences among solver-based platforms, rather than evaluating whether solver-based platforms outperform other existing routing designs.

\subsubsection{Binance Markout}

In addition to evaluating solver executed trade quality relative to AMMs, we also examine its comparison to the instantaneous midprice on centralized exchange order books. While acknowledging that the CEX midprice does not reflect the actual execution price for settling the same solver executed trade, it serves as a useful benchmark for prevailing market conditions. We calculate the CEX markout using the Binance data fetched from the Tardis API \cite{tardis2024} and the following formula:

\[
\text{CEX Markout} = \frac{\text{Solver Executed Price} - \text{Closest Binance Price}}{\text{Closest Binance Price}}
\]

where the closest Binance price is fetched as the next immediate available price in the dataset after the block timestamp of the trade executed by the solver. The total execution price is calculated as:

\[
\text{Solver Executed Price} = \frac{\text{Amount of Quote Asset}}{\text{Amount of Base Asset}}
\]

For buy trades, the sign of the CEX markout calculated from the above formula is flipped, as a lower price means better execution quality provided by the solver.

For example, if a solver executed a trade for USDC \(\rightarrow\) WETH happened on 2023-01-25 12:33 AM, and the next immediate data point in the ETH Binance price dataset is 2023-01-25 12:33:20 AM, then this data point is used to calculate the markout. If the solver-executed trade has \(1000\) USDC as input and \(1\) WETH as output, the solver executed price is:

\[
\frac{1000 \text{ USDC}}{1 \text{ WETH}} = 1000 \text{ USDC/WETH}
\]

If the Binance price is \(999\) USDC/WETH, then the CEX markout is calculated as:

\[
-1 \times \frac{1000 - 999}{999} = -0.1\%
\]

which means that the execution of the solver trade is \(0.1\%\) (10bps) worse than the CEX trade.

\subsubsection{Realized Volatility}

To test whether the solver's welfare is correlated with the overall market volatility, we calculated the realized volatility for the assets using the asset’s Binance price, in the 2-hour window before each solver executed trade. The formula is:

\[
\text{Realized Volatility} = \sqrt{\frac{1}{K} \sum_{t=1}^{K} \left( \log \frac{\text{prices}_{t+1}}{\text{prices}_t} \right)^2}
\]

where t is the block time when solver trades were executed and K is the total number of trades. For example, if a solver executed trade for WETH \(\rightarrow\) USDC happened on 2023-01-25 12:33 AM, the we find the corresponding ETH price on Binance from 2023-01-25 10:33 AM to 2023-01-25 12:33 AM and calculate the realized volatility for the solver executed trade using the above formula.

For analyzing the correlation between realized volatility and execution welfare, the weighted average execution welfare for the 2-hour window before each executed solver trade is calculated by normalizing the execution welfare by the trade size of each solver executed trade in the 2-hour window. This is given by:

\[
\text{Weighted Average Execution Welfare} = \frac{\sum_{i=1}^{K} \left( \text{Execution Welfare}_i \times \text{Trade Size}_i \right)}{\sum_{i=1}^{K} \text{Trade Size}_i}
\]

where \(K\) represents the number of solver-executed trades within the 2-hour window.

\section{Results}
In this section, we present the results of overall execution welfare across the three protocols and analyze its distribution across trade sizes. We then examine how this metric varies across different asset profiles. Furthermore, we discuss the differences in liquidity profiles across platforms and explore the correlation between execution welfare and volatility.
\subsection{Overall Execution Quality}
\begin{figure}[h]
    \centering
    \begin{minipage}{0.48\textwidth}
        \centering
        \includegraphics[width=\linewidth]{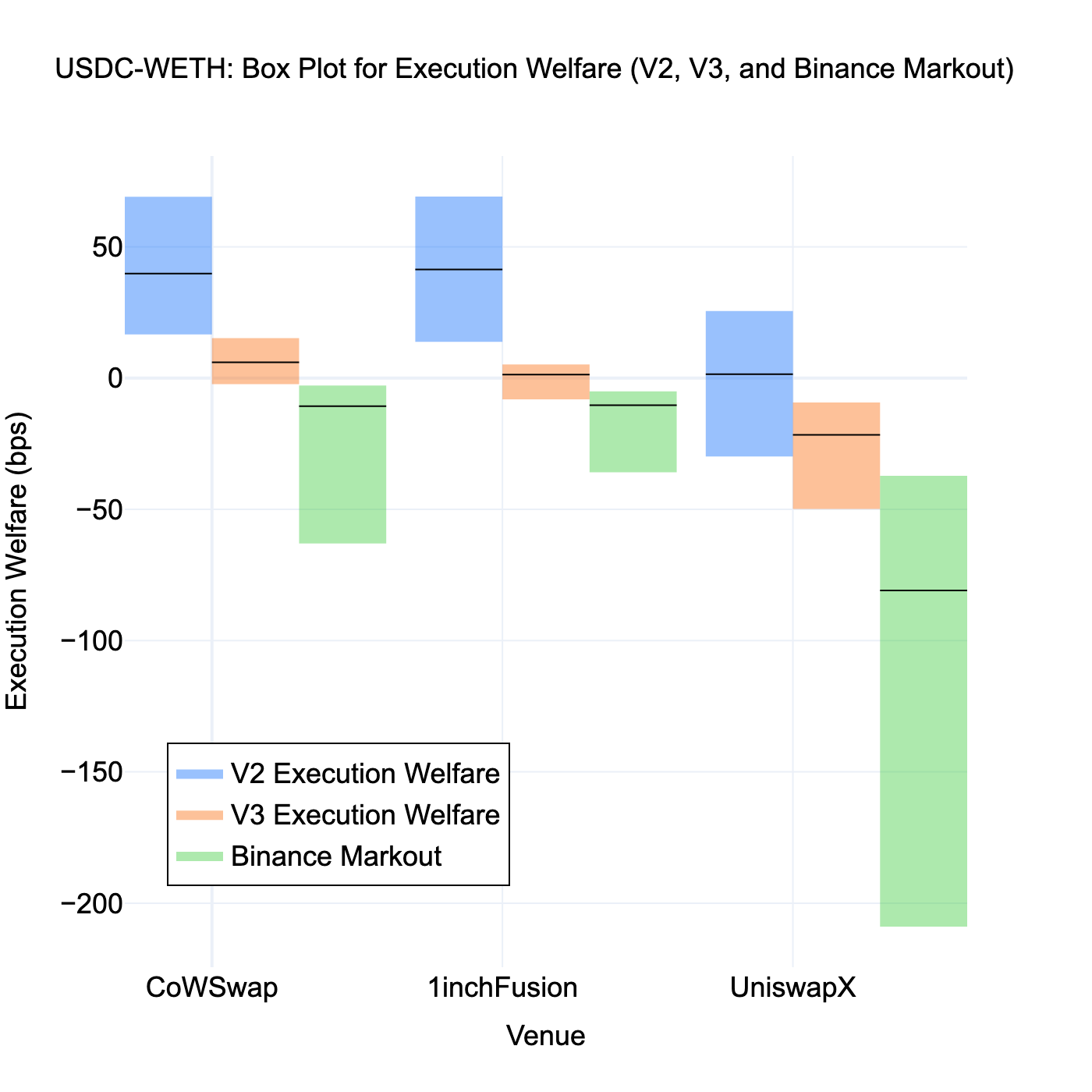}
        \caption{Boxplot of Welfare for USDC-WETH Trades}
        \label{fig:box-usdc}
    \end{minipage}
    \hfill
    \begin{minipage}{0.48\textwidth}
        \centering
        \includegraphics[width=\linewidth]{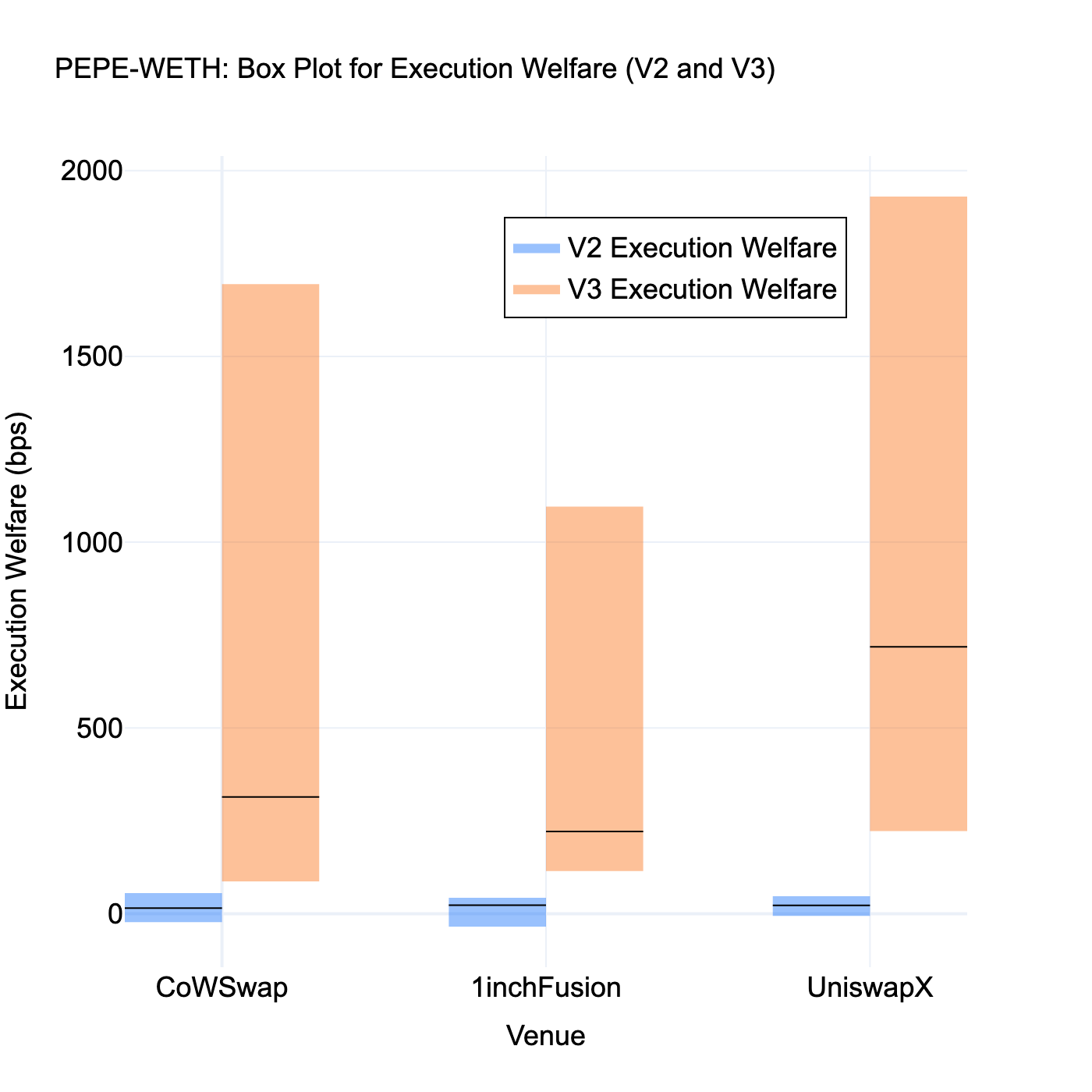}
        \caption{Boxplot of Welfare for PEPE-WETH Trades}
        \label{fig:box-pepe}
    \end{minipage}
\end{figure}

We present the overall execution welfare for USDC-WETH trades across the three venues using a box plot in Figure~\ref{fig:box-usdc}, which illustrates the median and quartile range (p25 and p75) of welfare values. The results indicate that execution welfare relative to V2 is consistently positive across all venues, demonstrating that solver-based routing provides a meaningful improvement over naive routing against V2 pools. In contrast, solver execution welfare relative to V3 remains marginally negative or close to zero, suggesting that V3’s concentrated liquidity model remains highly competitive and difficult to consistently outperform. 

Results show only CoWSwap consistently improves user welfare over V3 pools. The Binance markout reveals a negative trend, particularly for UniswapX, implying that the execution on this platform deviates more unfavorably from prevailing off-chain prices at the moment of trade.

Figure~\ref{fig:box-pepe} focuses on the quartile and median execution welfare of PEPE-WETH trades under the same evaluation framework. An important distinction in the PEPE-WETH dataset is the absence of markout data. This is due to the lack of Binance price data for the time duration analyzed in this study. 

In contrast to USDC-WETH, the improvement in execution welfare of the solver relative to V2 appears minimal across all venues. However, solver execution welfare relative to V3 is significantly positive, particularly for UniswapX and CoWSwap. Comparison of execution improvement across V2 and V3 suggests that V3 offers worse execution than V2, potentially due to the lack of V3 liquidity.

The results highlight that the trade execution improvement offered by solver-based DEXes is highly asset-dependent. For USDC-WETH, where market depth and liquidity fragmentation are high, solvers offer significant improvements over V2 and incremental improvement over V3. Meanwhile, for PEPE-WETH, where liquidity is mostly deposited into V2 pools and less fragmented, solvers significantly outperform V3 but not V2. This suggests that V3 and V2 remain efficient execution venues if highly concentrated liquidity is given. In contrast, in a market with fragmented liquidity, solver-based DEXes extract a greater surplus for traders.


\subsection{Execution Welfare over Trade Size}
To dive further into the behavior of execution welfare with varying trade sizes, for each type of asset pair, we plot the scatter plots for execution welfare over the increase of trade sizes. 

\begin{figure}
    \centering
    \includegraphics[width=1\linewidth]{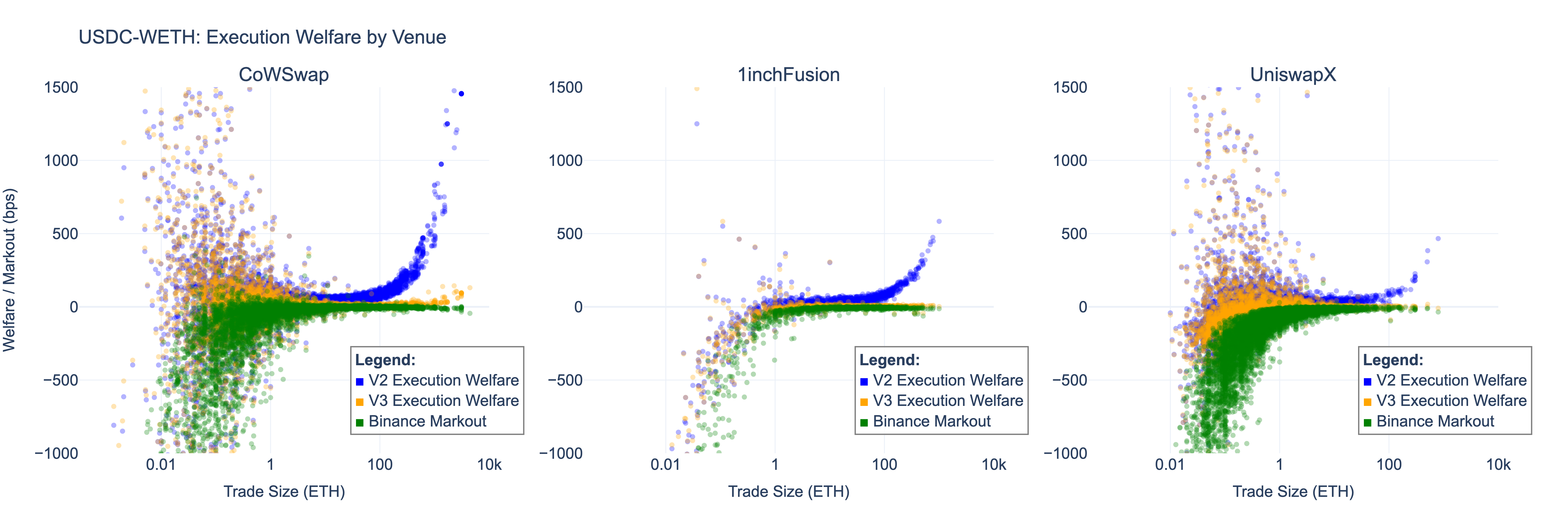}
    \caption{USDC-WETH Execution Welfare over Trade Size}
    \label{fig:welfare-usdc}
\end{figure}

\textbf{USDC-WETH Pair}
As depicted in Figure~\ref{fig:welfare-usdc}, for all three venues, trades over 100 ETH show an uptick in execution welfare compared to V2 going from 100bps and higher. This indicates that all three venues outperform V2 pools for large trades, providing favourable outcomes for end users when they swap large volumes. This pattern can be attributed to the rapidly increasing price impact on the V2 price curve as trade size increases.

CoWSwap, among the three, has the most pronounced improvement trend as trade volume increases - the V2 welfare stabilizes and achieves exponentially higher positive values of above 500 bps, indicating CoWSwap’s strength in delivering better trade outcomes for larger trades against V2. 

For both CoWSwap and UniswapX, V2 welfare shows significant variance in smaller trades, with pronounced outliers suggesting occasional highly favorable or negative trade outcomes relative to V2 and V3 pool simulations. However, for trade sizes smaller than 1 ETH, 1inchFusion shows a more consistent negative welfare.


CoWSwap's V3 welfare metric follows a similar trend but remains consistently lower than V2 welfare, indicating slightly reduced but better execution welfare compared to V3 pool. Toward the larger end of trade sizes, CoWSwap consistently provides over 20 bps welfare improvement with some large trades reaching near 150 bps. 

In the case of 1inchFusion, the V3 welfare remains lower than the V2 welfare by 10-50 bps but closely tracks its behavior in the small to medium-sized trades. For larger trades, 1inchFusion displays a moderate welfare improvement of around 5 to 10bps against V3. 

For UniswapX, V3 welfare shows a lower but positive value on larger trade sizes,  hovering around 0 to 10 bps welfare improvement. Data samples below the trade size of 5 ETH show a trend towards 0-to-negative welfare with decreasing trade size with high noise. 

The Binance markout across all three venues predominantly exhibits negative values for smaller trade volumes. However, as trade size increases from 0 ETH to 100 ETH, the negative markout narrows from below -1000 bps to a range of approximately -10 to -5 bps. This trend suggests that the relative impact of gas costs on on-chain execution diminishes with larger trades, resulting in prices that more closely align with prevailing market conditions reflected by the CEX midprice.

Overall, the three platforms demonstrate strong welfare performances for larger trades, consistently outperforming the counterfactual trades simulated on V2 pools. The results also showed that the welfare performance is less pronounced against V3 across the board. However, with increasing trade sizes, solver-based DEXes generally displayed positive execution welfare, with CoWSwap, in particular, presenting strong evidence of outperforming V2 and V3 liquidity. Lastly, as a reference to evaluate the alignment of on-chain execution with off-chain market conditions, all platforms'  execution welfare relative to Binance's markout remain negative, suggests that on-chain prices deviate unfavorably from prevailing off-chain prices at the moment of trade. However, this does not imply that Binance execution would necessarily be superior, as actual execution on Binance would incur slippage and impact, leading to prices worse than the midprice, particularly for large trades.

\begin{figure}
    \centering
    \includegraphics[width=1\linewidth]{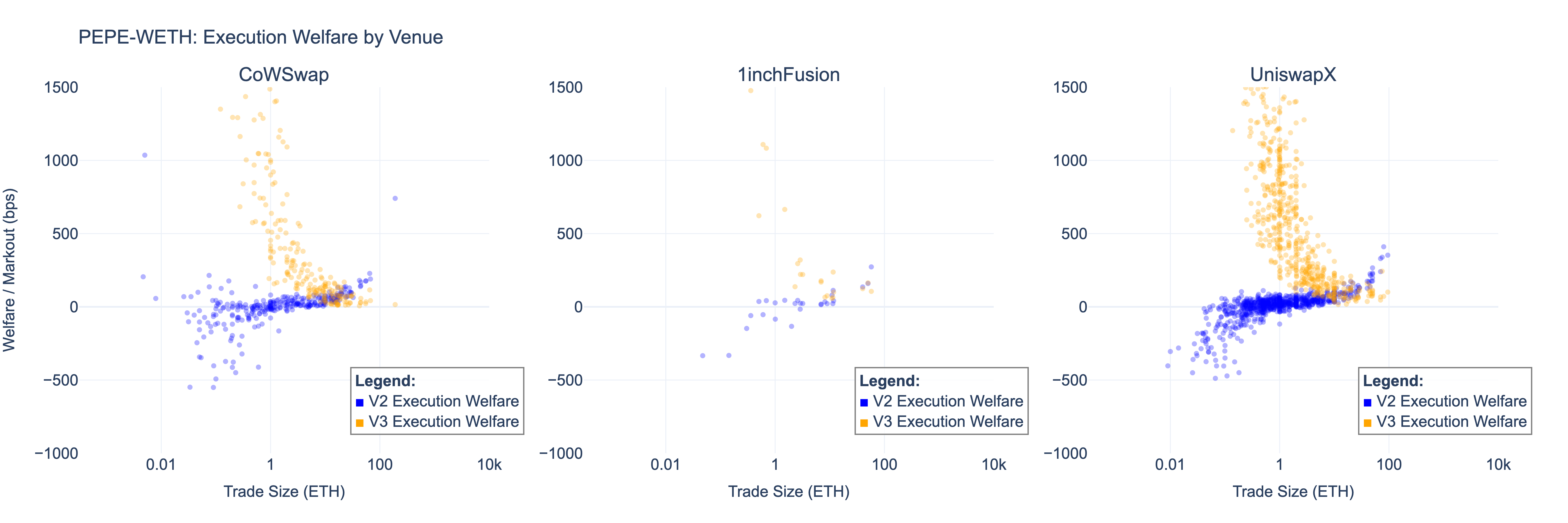}
    \caption{PEPE-WETH Execution Welfare over Trade Size}
    \label{fig:welfare-pepe}
\end{figure}

\textbf{PEPE-WETH Pair}
The scatter plots in Figure~\ref{fig:welfare-pepe} present the execution welfare for PEPE-WETH with varying trade sizes. During the period covered in this study, the PEPE-WETH V3 pool's liquidity declined from \$5 million to \$2 million while the V2 pool maintained a total liquidity of \$10 million. In contrast, the USDC-WETH pools had significantly deeper liquidity, with the V3 pool ranging from \$200 million to \$128 million, and the V2 pool holding approximately \$100 million. As a result, for all three venues, benchmarking against V3 pool exhibits a distinctly different shape compared to USDC-WETH.
Compared to counterfactual trades executed on V3, all CoWSwap trades exhibit significantly positive welfare, with trades smaller than 1 ETH exceeding 500 bps. As trade size increases, execution welfare gradually converges toward 0 bps, suggesting that execution quality declines to a level comparable to that of V3 pools. However, compared to V2, CoWSwap execution welfare exhibits an increase trend as trade size grows. This suggests that for large-size trade, its auction successfully reduced the pricing depreciation via batch auction and diverse liquidity, compared to direct AMM pricing curve of the V2 pool.

The welfare distribution for trades executed via 1inchFusion exhibits a similar dispersion pattern at smaller trade sizes. As trade size increases, execution welfare on 1inchFusion gradually converges toward 0 bps, indicating diminishing advantages over V3 execution. When compared to V2, execution welfare remains largely centered around 0 bps, with occasional increases for larger trades. However, the dataset is too sparse to draw conclusive insights on the overall welfare trend.

UniswapX exhibits a more consistent trend in execution welfare, particularly relative to V3. Many trades achieve substantial positive welfare, especially for small to moderate trade sizes, with several exceeding 500 bps. Compared to V2, while smaller trades show some negative welfare, larger trades display a more pronounced upward trend, indicating that UniswapX offers greater execution benefits at scale.

This pattern indicates that for long tail asset, solver-based DEXes performance against naive AMM routing varies depending on AMM pool's pricing model. When compared to a V3 pool, solver-based DEXes demonstrate superior execution, though with a diminishing advantage as trade size increases. In contrast, when compared to a V2 pool, solver-based DEXes offer improved outcomes over the increase of trade size, benefiting from a more gradual price impact growth.

Overall for PEPE-WETH trades of significant size, all solver-based DEXes facilitate more favorable trade outcomes relative to direct execution in either V2 or V3 liquidity pools.

\subsection{Liquidity Profile}

\begin{figure}[h]
    \centering
    \begin{minipage}{0.48\textwidth}
        \centering
        \includegraphics[width=\linewidth]{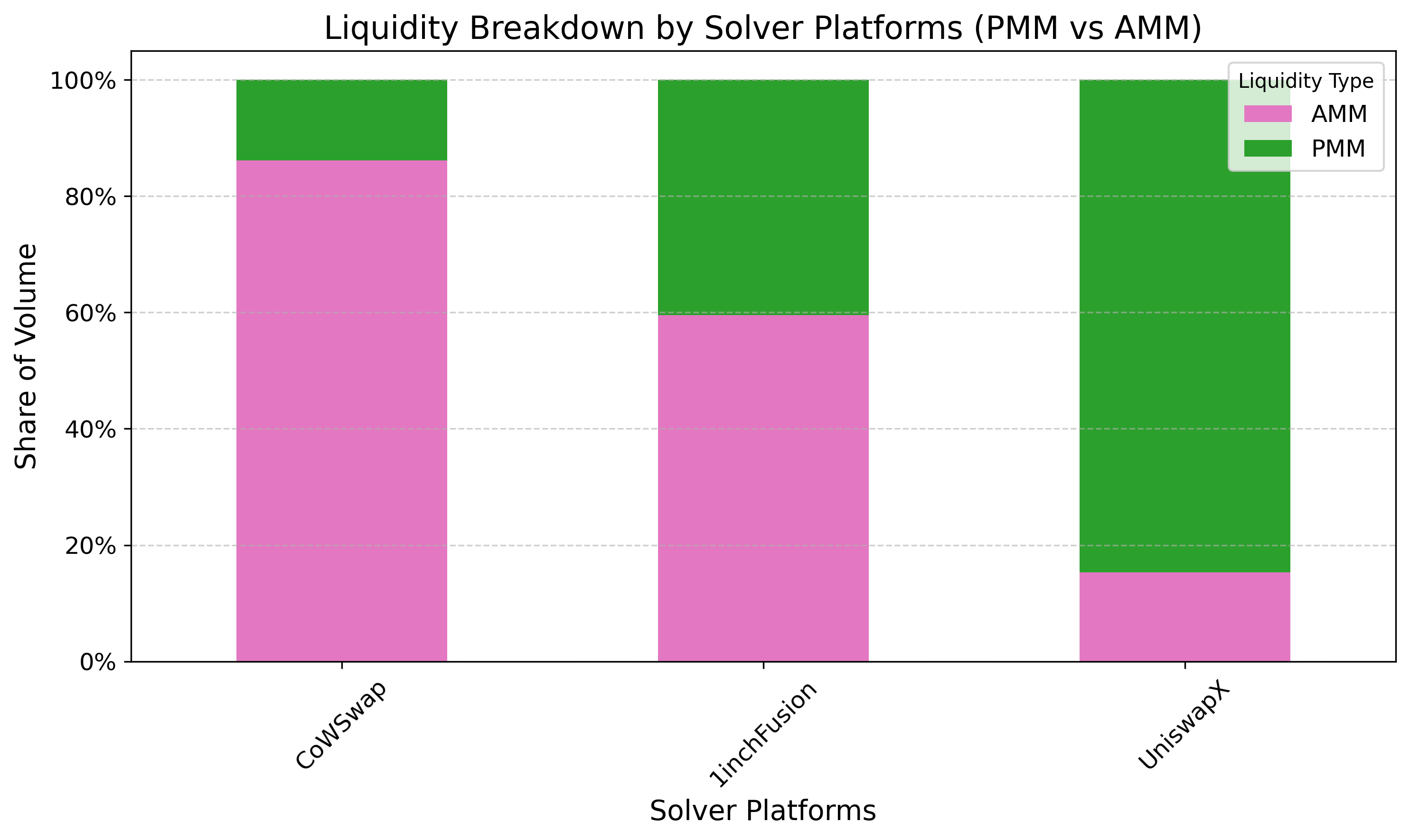}
        \caption{Liquidity Share Breakdown by Solver Platforms}
        \label{fig:liq-solver}
    \end{minipage}
    \hfill
    \begin{minipage}{0.48\textwidth}
        \centering
        \includegraphics[width=\linewidth]{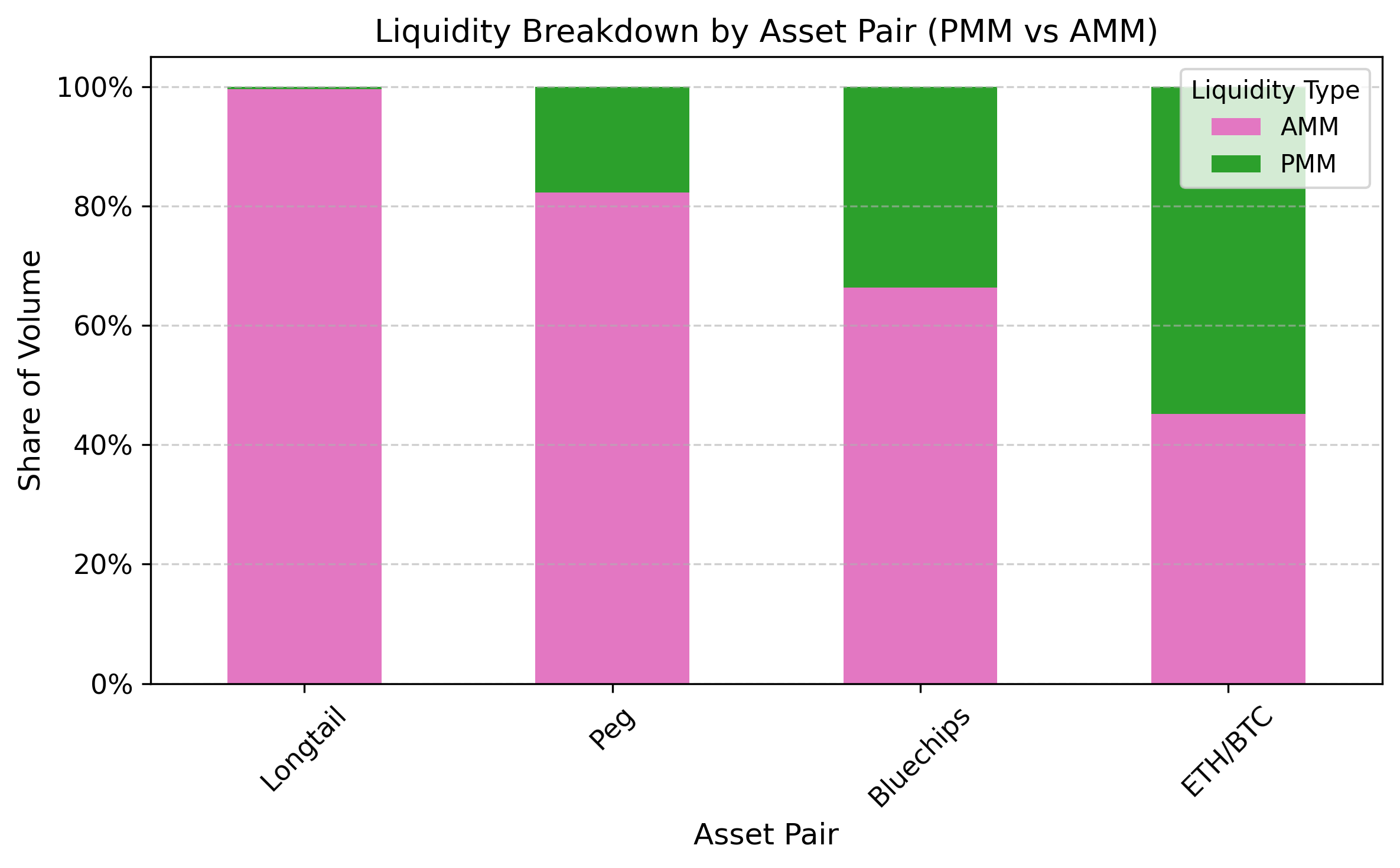}
        \caption{Liquidity Share Breakdown by Asset Pairs}
        \label{fig:liq-asset}
    \end{minipage}
\end{figure}

To further examine the factors contributing to variations in execution welfare across different platforms, we analyze the liquidity profile distribution across asset pairs and platforms. Despite similarities in solver auction workflows, the composition of solver profiles varies significantly across platforms in Figure~\ref{fig:liq-solver}. On CoWSwap, more than 85\% of executed trades are routed through AMM liquidity, whereas UniswapX exhibits the opposite distribution, with over 85\% of volume settled against PMM liquidity.

Figure~\ref{fig:liq-asset} shows that across all solver-based platforms, PMM participation increases as asset pairs become more prominent, with a general tendency to avoid extreme long-tail tokens. For example, in the case of PEPE-WETH—a memecoin listed on centralized exchanges—PMM are open to market make the pair as its inventory risk and execution costs has been lowered. While its profile leans toward the blue-chip token category, it remains a longer-tail asset relative to USDC-WETH.

\begin{figure}
    \centering
    \begin{minipage}{0.48\textwidth}  
        \centering
        \includegraphics[width=\linewidth]{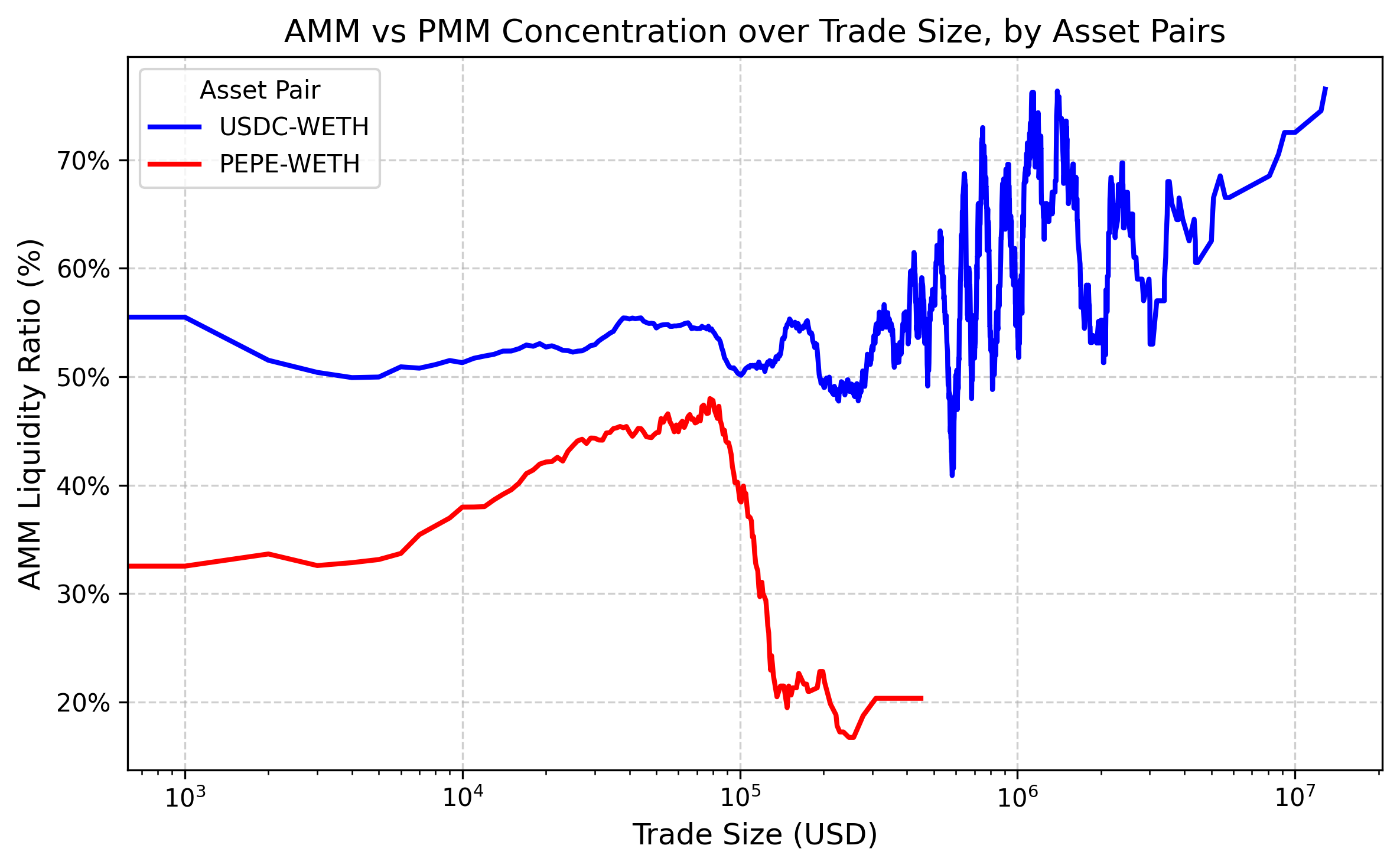}
        \caption{AMM Liquidity Concentration (MA50) by Trade Size (bin=1k)}
        \label{fig:ma50}
    \end{minipage}
    \hfill
    \begin{minipage}{0.48\textwidth}  
        \centering
        \includegraphics[width=\linewidth]{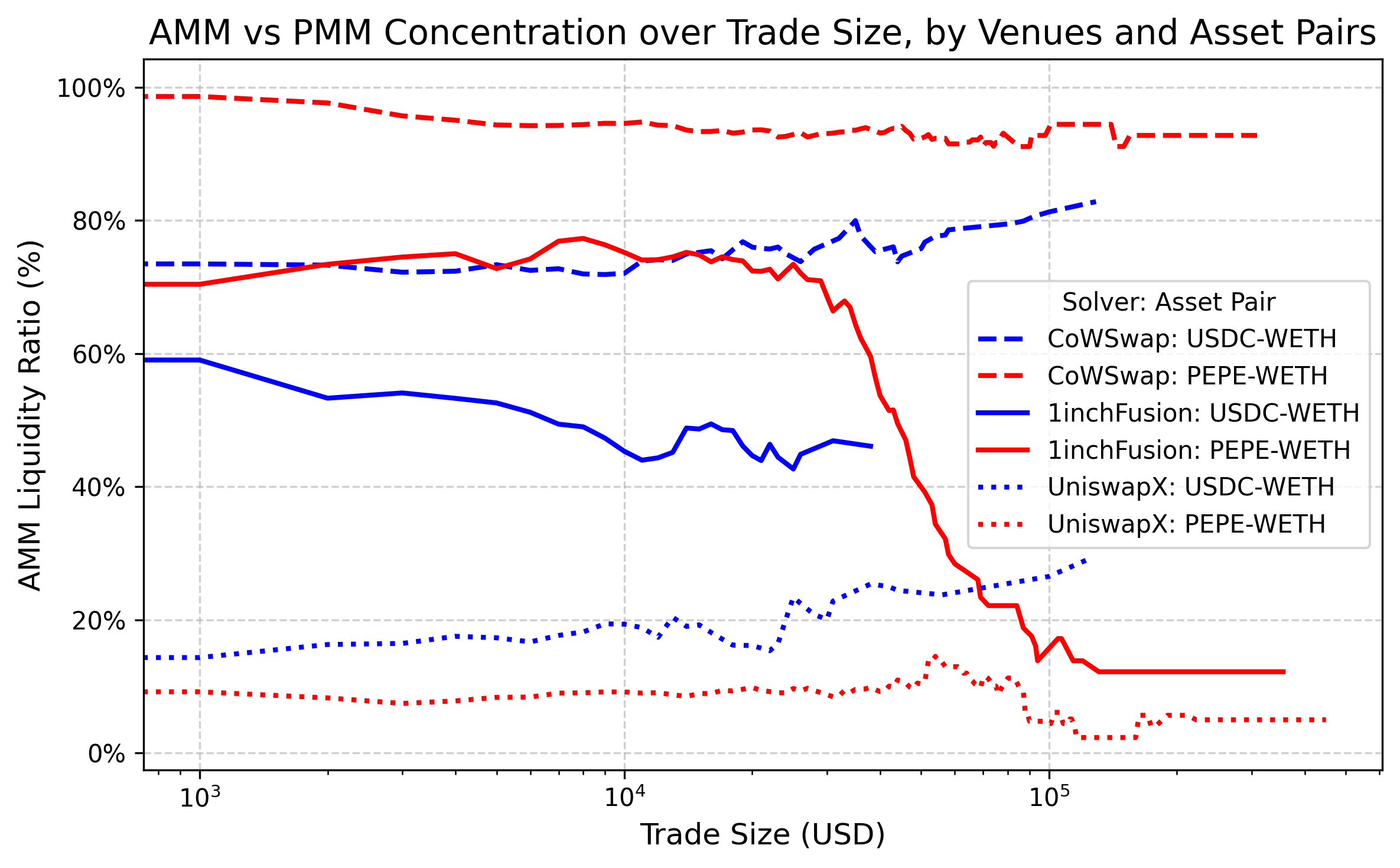}
        \caption{AMM Liquidity Concentration (MA30) by Trade Size (bin=1k)}
        \label{fig:ma30}
    \end{minipage}
\end{figure}

Furthermore, we analyzed the smoothed welfare using a moving average over trade size bucket windows ranging from 30 to 50 trades across different asset pairs and venues. 

For the USDC-WETH pair, combined across all solver-based venues, AMM and PMM liquidity exhibit an approximately even split for trade sizes below \$300,000 in Figure~\ref{fig:ma50}. However, for trade sizes exceeding \$300,000, AMM liquidity dominance shows an increasing trend.

In the case of the PEPE-WETH pair, AMM liquidity presence increases over time for trade sizes below \$75,000, rising from 32\% for smaller trades to nearly 50\%. However, once trade sizes exceed \$75,000, the percentage of trades settled via AMM liquidity declines rapidly. This trend suggests that for larger trades, PMMs holding PEPE achieve better execution compared to onchain liquidity, offering more superior pricing.

Since Figure~\ref{fig:ma50} is aggregated across venues, the results may be largely influenced by platforms with higher volume weights and data points. To better understand the nuances across different venues, we further breakdown the analysis by venue and asset pair. As shown in Figure\ref{fig:ma30}, each venue's AMM liquidity concentration remains consistent with its liquidity profile breakdown presented in Figure\ref{fig:liq-solver}, yet exhibits distinct trends across assets.

On CoWSwap, the PEPE-WETH pair is predominantly filled using AMM liquidity, though it shows a slight decline as trade size increases. Additionally, AMM liquidity maintains a higher presence in PEPE-WETH volume than the USDC-WETH pair on CoWSwap, aligning with the broader trend of liquidity dominance across asset pairs. However, on UniswapX, the PEPE-WETH pair exhibits a lower AMM liquidity ratio than the USDC-WETH pair, suggesting that the solver set on UniswapX is not risk-averse toward PEPE-WETH or may be strategically holding exotic asset inventories.

In contrast, 1inchFusion demonstrates a notable drop in AMM liquidity dominance for the PEPE-WETH pair as trade size approaches \$300,000. For the USDC-WETH pair, 1inchFusion solvers show a slight decline in AMM liquidity utilization, indicating that PMM liquidity is offering more competitive quotes as trade size increases.

\subsection{Volatility Factors}

We are curious whether execution welfare is influenced by market conditions and whether it improves or deteriorates during periods of volatility. To analyze the relationship between realized volatility and execution welfare, we plotted Figure  \ref{fig:vol-welfare-comparison} (see Appendix \ref{appendix:vol}) for the three venues. There appears to be no clear correlation between execution welfare and realized volatility for both Uniswap V2 (left) and Uniswap V3 (right). The data points are widely dispersed and do not follow any noticeable upward or downward trend, suggesting that changes in realized volatility have minimal to no consistent impact on execution welfare. 
After correlation analysis, we confirmed that there is no meaningful correlation between execution welfare (both V2 and V3) and realized volatility for the three venues.

\begin{table}[h]
    \centering
    \scriptsize
    \begin{tabular}{lcc}
        \hline
        & \textbf{corr(Volatility, V2 welfare)} & \textbf{corr(Volatility, V3 welfare)} \\
        \hline
        \textbf{CoWSwap}  & 0.0309 & 0.0267 \\
        \textbf{UniswapX} & 0.0497 & 0.0310 \\
        \textbf{1inchFusion}& 0.0464 & 0.0764 \\
        \hline
    \end{tabular}
    \caption{Correlation between Realized Volatility and Execution Welfare (V2 and V3)}
    \label{tab:volatility_welfare}
\end{table}


\section{Conclusion}
This study provides an empirical evaluation of execution welfare in solver-based decentralized exchanges (DEXes), analyzing their performance relative to traditional Automated Market Makers (AMMs) such as Uniswap V2 and V3. By leveraging onchain data and comparing execution quality across UniswapX, 1inchFusion, and CoWSwap, we present a structured assessment of how much price improvement solver-based designs bring for end users, on both short-tail (USDC-WETH) and long-tail (PEPE-WETH) asset pairs.

Our findings indicate that solver-based DEXes consistently offer better execution welfare compared to Uniswap V2, particularly for larger trade sizes where solvers effectively optimize routing strategies. However, when benchmarked against Uniswap V3, execution welfare improvements are less pronounced and vary based on asset liquidity. While solver-based execution generally outperforms V3 for long-tail assets like PEPE, the gains are marginal for more liquid pairs such as USDC-WETH, suggesting that Uniswap V3’s concentrated liquidity design remains highly competitive in execution efficiency.

A key insight from comparing the welfare of USDC-WETH, a short-tail asset, and PEPE-WETH, a long-tail asset, is the distinct execution welfare dynamics driven by differences in liquidity depth and type. For USDC-WETH, both V2 and V3 pool have relatively deep liquidity, which means the fragmentation can be solved by aggregation and routing by solvers. Solvers' execution welfare improves consistently with trade size across both V2 and V3, as deeper liquidity and more effective routing mechanisms enable better price execution at scale. 

In contrast, PEPE-WETH AMM pool liquidity is disproportionately weighted toward the V2 pool, with V2 holding five times more liquidity than V3. This imbalance is reflected in the welfare benchmark: this pair's execution welfare shows greater improvement at smaller trade sizes compared to V3, while larger trade sizes experience higher welfare gains against V2. This pattern is likely driven by the concentration of PEPE-WETH liquidity in V2 rather than V3, causing most trades to default to V2 routing. As a result, the benefits of routing and liquidity aggregation are limited, particularly for smaller trades that are already efficiently executed within the existing V2 liquidity. For larger trades, however, solver-based venues gain a greater advantage by integrating offchain liquidity and optimizing order execution, leading to increased execution welfare relative to V2. This contrast underscores how asset-specific liquidity distributions shape the effectiveness of solver-based routing and execution strategies across different trading venues.

Additionally, our results show that solver-based execution tends to produce negative markouts relative to the Binance midprice across all cases. This suggests that despite solvers’ ability to aggregate on-chain liquidity and utilize RFQ mechanisms, on-chain execution prices often deviate unfavorably from prevailing market conditions, particularly for small to medium trade sizes.

By examining liquidity profile across solver-based platform, we highlighted the varying onchain versus offchain liquidity dominance across solver set. Furthermore, the analysis of realized volatility found no significant correlation between execution welfare and market fluctuations. This suggests that the execution quality of solver-based DEXes is primarily driven by microstructure and factors such as trade size, liquidity fragmentation, and solver competition, rather than broader market conditions.

\section{Limitations and Future Work}




While our study provides valuable insights, several limitations and underlying assumptions should be acknowledged:
\begin{itemize}
    \item \textbf{Limited Asset Coverage} – USDC-WETH and PEPE-WETH pairs were selected to represent short- and long-tail asset profiles. PEPE-WETH was chosen to ensure a sufficient trade volume for a meaningful data sample. However, due to its adoption on centralized exchanges, PEPE-WETH no longer fully represents a long-tail asset. Expanding the analysis to include additional asset pairs across different categories could provide more robust results.
    \item \textbf{Liquidity Snapshot Assumption} – The execution welfare calculation is based on counterfactual simulations using Uniswap V2 and V3 liquidity snapshots at the top of each block. However, this approach does not capture intra-block liquidity distribution changes. Factors such as trade collisions and Maximal Extractable Value (MEV) activity can impact price movements in real-world conditions, affecting the execution environment for solver-executed trades.
    \item \textbf{Simulation without Routing} -  Because the simulation tool does not route across multiple pools, the simulated AMM price does not fully reflect the optimal routing outcomes achievable by modern aggregator routers, which can split and distribute trade sizes across various AMM pools and protocols. As a result, our benchmarks only measure price improvement relative to a single Uniswap AMM pool. Future analysis could incorporate open-source routing algorithms to evaluate how solver-based platforms compare with modern DEX aggregators that utilize advanced routing and PMM liquidity via RFQ. This would allow for a more precise assessment of the price improvement attributable specifically to the solver auction mechanism.
    \item \textbf{Selection Bias on Filled Orders} – Our dataset includes only executed trades that landed onchain, without considering unfilled orders. This may introduce selection bias, particularly if solvers selectively execute only profitable trades. Our execution welfare metric does not capture the potential adverse selection behavior from sophisticated solvers.
    \item \textbf{Batch Auction Delay} - Another aspect of execution quality not captured by the welfare metric is the time delay from the user's perspective. Batch auction designs, such as CoWSwap, may introduce longer delays as the platform waits for additional orders before execution. Orders filled significantly later in a volatile market effectively equals to a worse price, representing an implicit cost to the user. By simulating AMM prices at the top of the block in which trades are executed, the analysis evaluates execution quality based on block-time execution rather than the actual time the user submitted the order. Since order request timestamps are not recorded on-chain, accurately estimating this delay remains challenging.


    \item \textbf{Gas Cost Estimation} – The estimation of gas costs for counterfactual AMM simulations assumes a conservative priority fee-adjusted gas price. While this provides a realistic estimate, small adjustment to the estimation may change the comparison dramatically. 
    \item \textbf{Solver Reward Incentives} – It is also worth noting that there is one major factor that contributes to the execution welfare whose effect is difficult to quantify per trade, namely token incentive. Solvers typically receive token rewards based on successfully filled order volume and price improvement. For example, CoWSwap gave 16 million \$COW~\cite{cow2024rewards} and 1inchFusion gave 10 million \$1INCH distributed to solvers annually~\cite{1inch2023twitter}, which help offset their gas and solving costs and subsidize less profitable solvings that, in turn, provide better trade executions for users. 



\end{itemize}

\section{Acknowledgement}
This study was funded by the TLDR fellowship grant. Special thanks to Kshitij Kulkarni for the extensive amount of mentorship, Ryan Shan for contributions to the dataset, 
and John Beecher, Andrea Canidio, Burak Öz, Julian Ma, Xin Wan, Max Resnick and Mallesh Pai for their feedback on early drafts.


\bibliographystyle{unsrt}     
\bibliography{refs}

\appendix
\section{Data Collection: Executed Trades} \label{appendix:data-executed}
\subsection{UniswapX Trades}

Etherscan is used to retrieve all transactions that call the UniswapX settlement contract (deployed at \texttt{0x6000...45C4})~\cite{etherscanUniswapX} on the Ethereum mainnet. Only transactions with a function signature \texttt{0x3f62192e}, which matches the \texttt{Execute()} function call, are collected. Each of these function calls represents when a solver fills a user's order being filled by a solver. The UniswapX SDK~\cite{uniswap2024sdk} is used to parse the calldata of each \texttt{Execute()} function call to extract the maker address, recipient address, input asset, and output asset. Orders with more than 1 output asset are discarded from the analysis, as the focus is solely on the USDC-WETH and PEPE-WETH trading pairs. Only trades where the input assets and output assets are WETH, USDC, or PEPE are included in the analysis.

UniswapX does not automate the process of wrapping and unwrapping native ETH, and it allows users to specify whether native ETH or WETH is desired as the output asset. Therefore, a special procedure is implemented in the data collection script to incorporate transactions where the output asset is native ETH. Specifically, if the output asset is \texttt{0xeeee...eeee} or \texttt{0x0000...0000}, the transaction is included in the analysis, with the respective output asset field being replaced with \texttt{0xc02a...6cc2}~\cite{etherscanWETH}, which is the contract address for WETH. If the input and output assets are ERC-20 tokens such as WETH, USDC, or PEPE, the input amount and output amount are extracted from the ERC-20 transfer events of the transaction. Alternatively, if the output asset is native ETH, the output amount is obtained from the internal transaction list of the transaction, retrieved from Etherscan.

\subsection{CoWSwap Trades}
CoWSwap settlement contract, deployed at \texttt{0x9008...ab41}~\cite{etherscanCoWSwap} on the Ethereum mainnet, emits a \texttt{Trade()} event for each settlement call. Infura is used to fetch event emissions by the CoWSwap settlement contract. The event has a data field \texttt{"orderID"}, which can be used to fetch the order details from CoWSwap API endpoint~\cite{cow2024orderbook}. The fetched details include the address for the maker, recipient, input asset, output asset, input amount, and output amount. Only trades where the input assets and output assets are WETH, USDC, or PEPE are included in the analysis. The taker address is fetched via the transaction's \texttt{Settlement()} event emission. CoWSwap does not automatically wrap/unwrap native ETH, thus a process similar to how we handled UniswapX is implemented, which regards \texttt{0xeeee...eeeee} and \texttt{0x0000...0000} as \texttt{0xc02a...6cc2}~\cite{etherscanWETH}.

\subsection{1inchFusion Trades}
Etherscan is used to retrieve all transactions that call the 1inchFusion settlement contract (deployed at \texttt{0xA888...7647})~\cite{etherscan1inchFusion} on the Ethereum mainnet. Only transactions with the function signature \texttt{0x0965d04b}, which matches the \texttt{SettleOrder()} function call, are collected. Each of these function calls represents a user's order being filled by a solver. The addresses for maker, taker, and recipient, as well as the addresses for input and output assets, and the amount of input asset, can be extracted from the calldata of each \texttt{SettleOrder()} transaction. Only trades where the input assets and output assets are WETH, USDC, or PEPE are included in the analysis. If the recipient field address is \texttt{0x0000...0000}, the recipient address is then set to the maker address.

1inchFusion automatically wraps and unwraps WETH, and all swaps are facilitated with WETH, irrespective of whether the specified output asset is native ETH. Therefore, the output amount is calculated by retrieving all ERC-20 transfer event emissions from the transaction, totaling the ERC-20 balance changes by address and asset, and obtaining the net changes of the output asset for the recipient address.

\section{Gas Estimation}\label{appendix:gas}

\begin{figure}[h]
    \centering
    \begin{minipage}{0.48\textwidth}
        \centering
        \includegraphics[width=\linewidth]{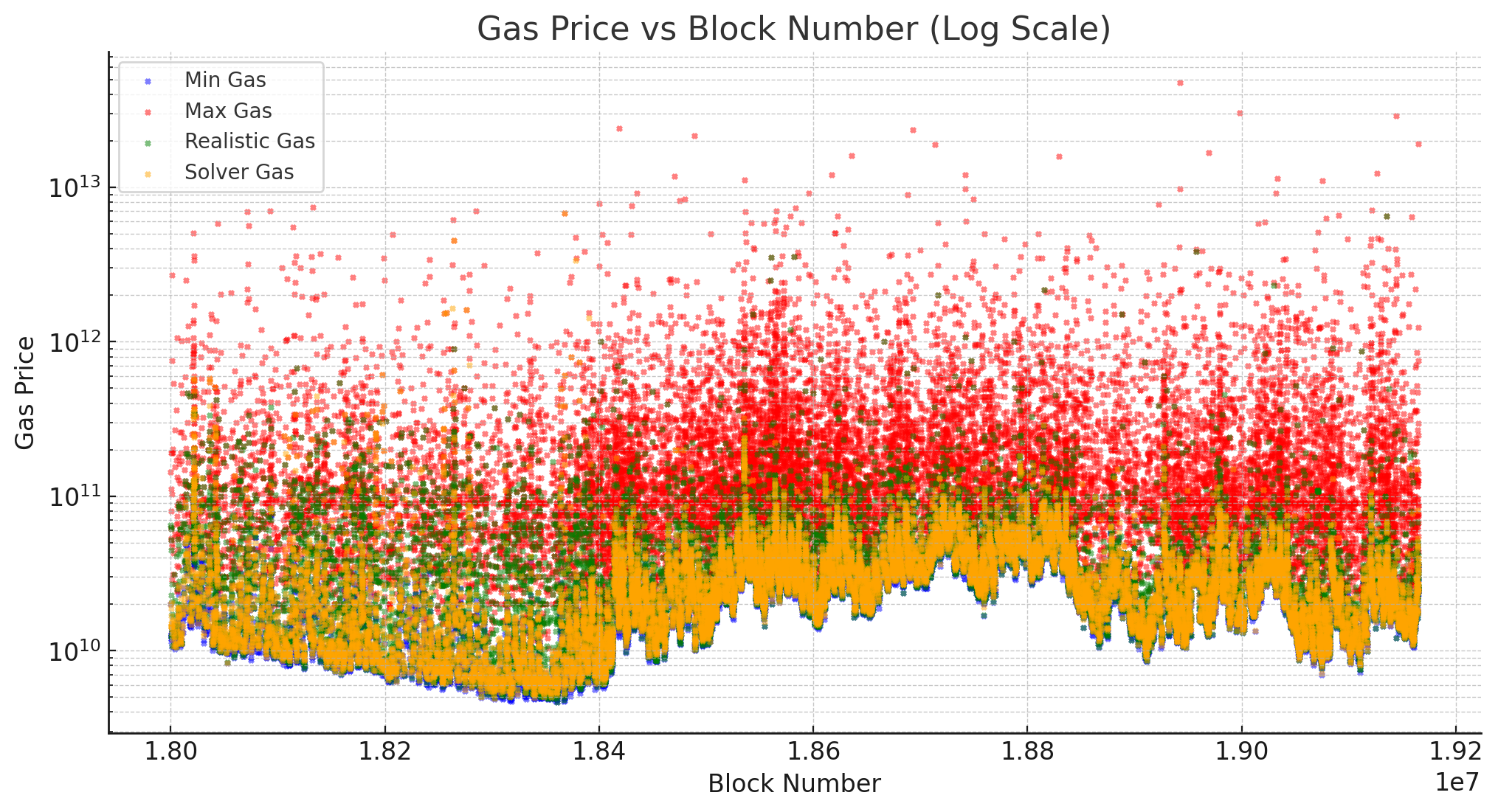}
        \caption{Gas price comparison across blocks}
        \label{fig:all-gas}
    \end{minipage}
    \hfill
    \begin{minipage}{0.48\textwidth}
        \centering
        \includegraphics[width=\linewidth]{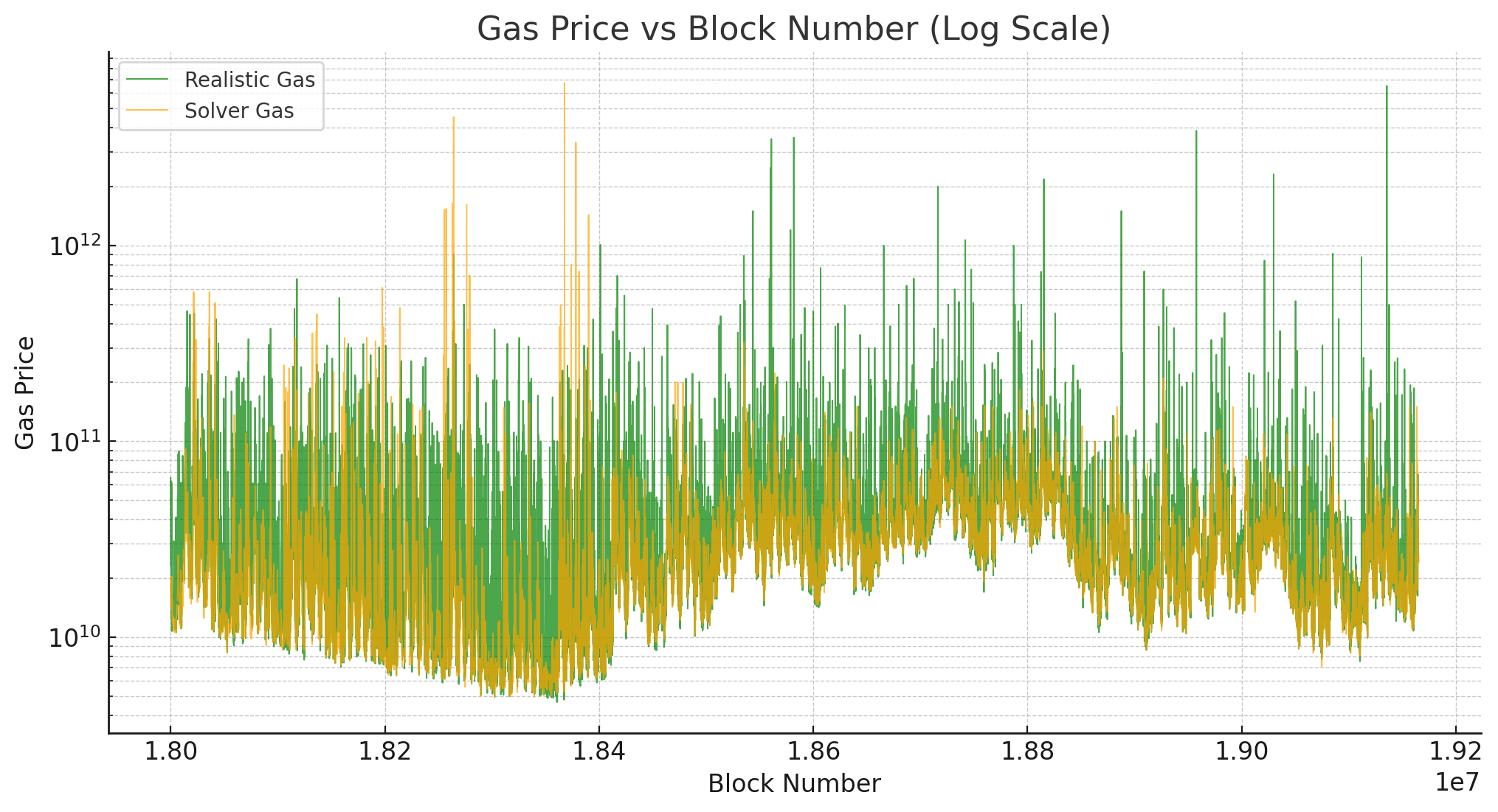}
        \caption{Solver gas price VS priority fee adjusted gas price (realistic)}
        \label{fig:solver-gas}
    \end{minipage}
\end{figure}
We studied the gas price distribution of the covered data range. Figure \ref{fig:all-gas} shows the distribution of the maximum gas price (red), minimum gas price (blue), the gas price paid for the first swap across the AMMs in the corresponding block (green, referenced as \textit{Realistic Gas}) and the gas price paid by the solvers (orange). Figure\ref{fig:solver-gas} extracts and compares only the solver gas and gas price paid for the first swap across AMMs. These results indicate that the highest gas price paid for swaps, in green, is closely mapped over the solver gas price, in orange. The average values for green and orange are \(41.7\) Gwei and \(32.5\) Gwei correspondingly. Given the higher average, we used the former gas price as our conservative estimate for simulated swaps. To summarize, our analysis will comprise of two approaches for gas estimation:

\begin{itemize}
    \item Under the \textbf{optimistic approach}, we fetch all transactions in the same block of the fill order transaction of the solver and calculate the minimum gas price observed in the corresponding block. Optimistic gas price provides insight on the best possible execution of the simulated swaps with little to no contention for the blockspace.
    
    \item Under the \textbf{priority fee adjusted (conservative) approach}, a full list of transactions in the same block is obtained, and the gas price paid by the first transaction in the block interacting with AMM DEXes (e.g. SushiSwap, Uniswap) or Aggregators (e.g. 1inch) is used as the gas price for the simulation process. If there is no transaction interacting with AMM DEXes or Aggregators in the same block of the solver fill order transaction, the gas price paid by the solver fill order transaction is used for the simulation process. This approach shows a conservative execution of the simulated swaps with a more realistic blockspace contention.
\end{itemize}

The optimistic gas cost is then estimated with:

\[
\text{Optimistic Gas Cost} = \text{Gas Units} \times \text{Optimistic Gas Price}
\]

Similarly, the conservative gas cost is estimated with:

\[
\text{Conservative Gas Cost} = \text{Gas Units} \times \text{Conservative Gas Price}
\]

for both Uniswap V2 and Uniswap V3 simulations.

\section{Volatility Factors (USDC-WETH)}\label{appendix:vol}
For each venue, the x-axis represents the weighted average (weighted by trading volume) execution welfare calculated over a 2-hour window before each solver-executed trade, while the y-axis shows the realized volatility during the same time window. Each point in the scatterplot corresponds to a solver-filled order, with its position indicating the execution welfare and realized volatility associated with that trade. Different colors (blue for CoWSwap, orange for UniswapX, and red for 1inch) were used to distinguish between the datasets. This visualization provides an intuitive way to assess potential correlations between the two variables across different venues.
\begin{figure}[h]
    \centering
    \begin{minipage}{0.48\textwidth}
        \centering
        \includegraphics[width=\linewidth]{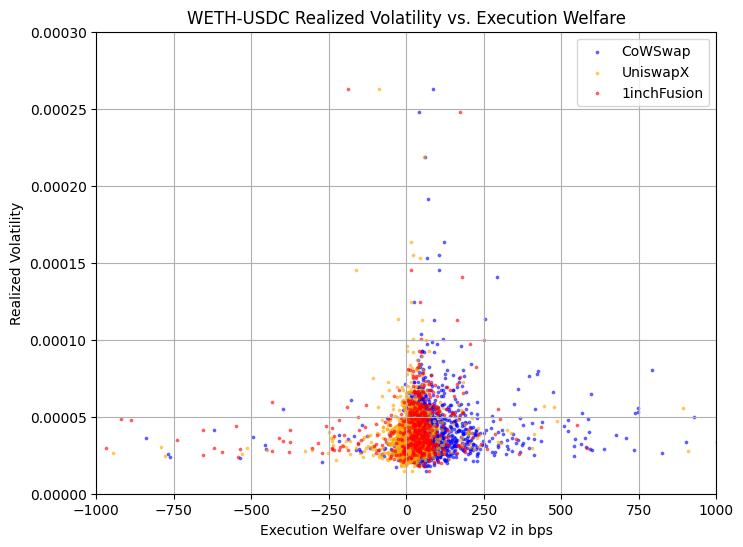}
    \end{minipage}
    \hfill
    \begin{minipage}{0.48\textwidth}
        \centering
        \includegraphics[width=\linewidth]{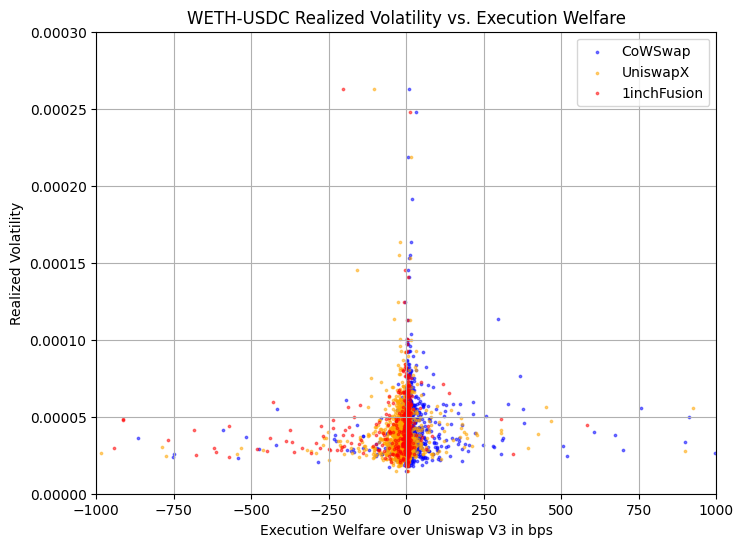}
    \end{minipage}
    \caption{Realized Volatility vs. Execution Welfare over Uniswap V2\textbf{ (left) }and V3 \textbf{(right)}}
    \label{fig:vol-welfare-comparison}
\end{figure}

\end{document}

%% file: defs.tex
\newcommand{\BEAS}{\begin{eqnarray*}}
\newcommand{\EEAS}{\end{eqnarray*}}
\newcommand{\BEA}{\begin{eqnarray}}
\newcommand{\EEA}{\end{eqnarray}}
\newcommand{\BEQ}{\begin{equation}}
\newcommand{\EEQ}{\end{equation}}
\newcommand{\BIT}{\begin{itemize}}
\newcommand{\EIT}{\end{itemize}}